\documentclass[12pt]{iopart}

\usepackage{iopams}  
\usepackage{graphicx}

\newcommand{\D}{\mathrm{d}}
\newcommand{\Bb}{\textbf{B}}
\newcommand{\Eb}{\textbf{E}}
\newcommand{\vb}{\textbf{v}}
\renewcommand{\sb}{\textbf{s}}
\newcommand{\Omegab}{\boldsymbol{\Omega}}

\begin{document}

\title[On the robustness of spin polarization for MVA protons in density down-ramps]{On the robustness of spin polarization for magnetic vortex accelerated proton bunches in density down-ramps}

\author{L Reichwein$^1$, A H\"{u}tzen$^{2,3}$, M B\"{u}scher$^{2,3}$, A Pukhov$^1$}
\address{$^1$Institut f\"{u}r Theoretische Physik I, Heinrich-Heine-Universit\"{a}t D\"{u}sseldorf, 40225 D\"{u}sseldorf, Germany}
\address{$^2$Peter Gr\"{u}nberg Institut (PGI-6), Forschungszentrum J\"{u}lich, 52425 J\"{u}lich, Germany}
\address{$^3$Institut f\"{u}r Laser- und Plasmaphysik, Heinrich-Heine-Universit\"{a}t D\"{u}sseldorf, 40225 D\"{u}sseldorf, Germany}
\ead{lars.reichwein@hhu.de}
\vspace{10pt}
\begin{indented}
\item[] \today
\end{indented}

\begin{abstract}
We investigate the effect of density down-ramps on the acceleration of ions via Magnetic Vortex Acceleration (MVA) in a near-critical density gas target by means of particle-in-cell simulations. The spin-polarization of the accelerated protons is robust for a variety of ramp lengths at around 80\%. Significant increase of the ramp length is accompanied by collimation of low-polarization protons into the final beam and large transverse spread of the highly polarized protons with respect to the direction of laser propagation.
\end{abstract}

%
\vspace{2pc}

%
\noindent Accepted for publication in \textit{Plasma Phys. Control. Fusion}
%
%
%

	\section{Introduction}
The acceleration of spin-polarized particles is interesting for a variety of applications, from testing the Standard Model of particle physics \cite{Androic2018} to examining the structure of subatomic particles for further insight on QCD \cite{Burkardt2010}. As laser-plasma based acceleration mechanisms have grown to be more prominent due to the high achievable energies over a shorter distance than in conventional accelerators \cite{Pukhov2002, Faure2004}, it is the logical next step to study the acceleration of spin-polarized particles in these regimes. The current state-of-the-art is given in the paper by B\"{u}scher et al. \cite{Buescher2020}.

In the case of  electrons, Wu et al. \cite{Wu2019, Wu2019a} have shown that via both laser-driven and particle beam-driven wakefield acceleration, high degrees of polarization can be achieved, if an appropriately chosen laser pulse or driving beam, respectively, are used. It could be seen that the real crux for generating high-polarization electrons lies within the injection: due to strong azimuthal magnetic fields during injection, the spins of the electrons start to precess strongly, leading to a significant loss of polarization, while during the acceleration phase, changes in polarization are mostly negligible.

For the acceleration of protons in general, various methods like Target Normal Sheath Acceleration (TNSA) \cite{Roth2016}, Radiation Pressure Acceleration (RPA) \cite{Macchi2017} or Magnetic Vortex Acceleration (MVA) \cite{Nakamura2010, Willingale2006, Willingale2008} are feasible options. 
Wakefield acceleration of protons is also possible, although significantly higher laser intensities are necessary \cite{Shen2007, Huetzen2019}. 
If we, however, need spin-polarized beams, we have to restrict ourselves to setups where we can pre-polarize our targets, ruling out some of the options due to the properties of the materials that are needed. Pre-polarizing the particles to be accelerated is necessary, since at the time scales and field strengths considered for acceleration, significant polarization build-up during the process is not possible \cite{Thomas2020}.

Jin et al. \cite{Jin2020} recently considered the acceleration of spin-polarized protons using a near-critical density target.
The process, identified as MVA, works as follows:
When the laser pulse enters the target, the ponderomotive force pushes the electron in the direction transverse to laser propagation, leaving behind a channel of low electron density \cite{Park2019, Nakamura2010}. Electrons can be accelerated in the wake induced by the laser and form a central filament in the channel. A strong azimuthal magnetic field is created by a current flowing in the central filament along the axis and an opposing current along the channel wall. This current also accelerates some ions in the filament structure along the channel center. When leaving the interaction volume, the magnetic fields can expand in the transverse region because of the decrease in density. Strong longitudinal and transverse electric fields are induced by the displacement of the electrons with respect to the ions. Finally, an ion beam is obtained that is further accelerated by the prominent fields after leaving the plasma.
Jin et al. showed that while higher intensities lead to better energies ($\mathcal{E}_p >  100$ MeV for a laser with normalized laser vector potential $a_0 = e A_0 / (m_e c) =100$), it comes at the price of lower polarization.
Here, $m_e$ denotes the electron mass and $c$ the vacuum speed of light.

In this paper, we investigate the effect of density down-ramps at the end of the interaction volume onto the obtained proton bunches, specifically the degree of polarization. We consider a gaseous HCl target similar to Ref. \cite{Jin2020} in our PIC simulations, keeping all parameters except the length of the down-ramp fixed throughout. 
It is shown that the degree of polarization is robust against down-ramp length and that obtaining high-quality bunches is mainly limited by the change in spatial beam structure due to the prevalent electromagnetic fields. Only for longer ramps the spatial structure is modulated so strongly, that lower-polarization protons are collimated into the beam. The results presented are discussed in the scope of the scaling laws of Ref. \cite{Thomas2020}.
 We find that the accelerated proton bunch can be described as consisting of three components, namely its back, middle and front. These parts contain protons from different states of the acceleration process, leading to distinct average polarizations. The extent of each of those parts is determined by the slope of the down-ramp that influences the focusing of the protons into the bunch and in turn the final beam quality.

\begin{figure}[t]
	\includegraphics[width=\textwidth]{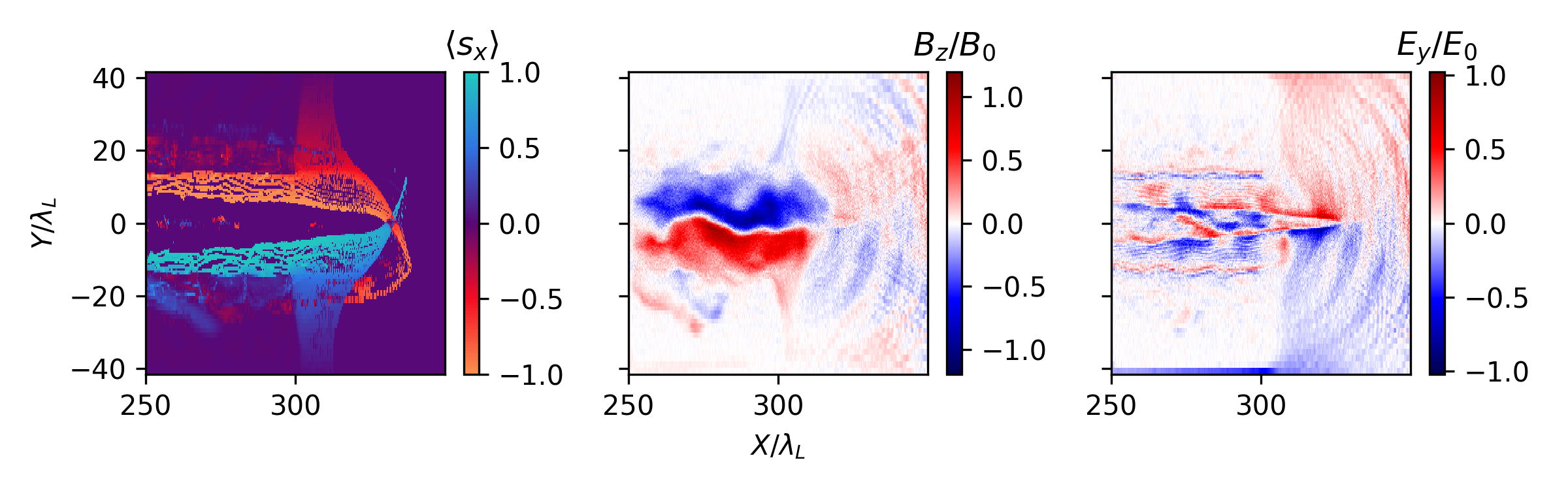}
	\caption{\label{fig:emfields}Distribution of particle spin and field configuration for the case of $L_\mathrm{ramp} = 0 \lambda_L$ at $t = 320 \tau_0$. All protons in the plasma have initial polarization $s_y = 1$. The electromagnetic fields are normalized with $E_0 = B_0 = m c \omega_{0} / e$. It can be seen that the accelerated proton bunch leaving the plasma maintains a high degree of polarization, while protons surrounding the remaining filament of the coaxial channel gain transverse polarization.}
\end{figure}

\section{Simulation setup}
For our simulations we use the PIC code VLPL \cite{Pukhov1999} that includes the precession of particle spin $\sb$ according to the T-BMT equation
\begin{equation}
	\frac{\D \sb_i}{\D t} = - \Omegab \times \sb_i \; ,
\end{equation}
where
\begin{equation}
	\Omegab = \frac{q}{mc} \left[ \Omega_\Bb  \Bb - \Omega_\vb \left( \frac{\vb}{c} \cdot \Bb \right) \frac{\vb}{c} - \Omega_\Eb \frac{\vb}{c} \times \Eb \right]  \label{eq:prec}
\end{equation}
is the precession frequency for a particle with charge $q$, mass $m$ and velocity $\vb$. The prefactors are given as
\begin{equation}
	\Omega_\Bb = a + \frac{1}{\gamma} \; ,  \Omega_\vb = \frac{a \gamma}{\gamma + 1} \; ,   \Omega_\Eb = a + \frac{1}{1 + \gamma} \; , \label{eq:pref}
\end{equation}
with $a$ and $\gamma$ being the particle's anomalous magnetic moment and its Lorentz factor, respectively.
This equation describes the change in spin for a particle that traverses through some arbritary configuration of electric fields $\Eb$ and magnetic fields $\Bb$.

In general, more spin-related effects would have to be considered, like the Stern-Gerlach force that describes the effect of spin onto a particle's trajectory, and also the Sokolov-Ternov effect, that links radiative effects with spin.
It has, however, been shown by Thomas et al. \cite{Thomas2020}, that these two effects can be neglected for the parameter regimes considered in the following. 

For our setup, we choose a circularly polarized laser with $a_0 = 25$ and wavelength $\lambda_L = 800$ nm. The length of the pulse is $\tau_0 = 10 \lambda_L / c$ and it has a focal radius of $w_0 = 10 \lambda_L$ (at $1/e^2$ of the intensity). 

The target consists of HCl gas with a peak density of $n_\mathrm{H} = n_\mathrm{Cl} = 0.0122 n_\mathrm{cr}$, leading to a near-critical electron background. Here, $n_\mathrm{cr}$ denotes the critical density. This specific gas is chosen because it allows for an easily achievable pre-polarization of the protons (see Ref. \cite{Wu2019a} for a detailed description of the process).
In our case, for all protons, we initially choose $s_y = 1$.

The interaction volume starts with an up-ramp rising from vacuum to peak density over a distance of $5 \lambda_L$, then maintaining peak density for $200 \lambda_L$. The down-ramp length at the end is varied in the range of $0\lambda_L$ up to $100 \lambda_L$ (see Table \ref{tab:results}).

In our simulations, we use a box of size $(100 \times 60 \times 60) \lambda_L$ that is moving alongside the laser pulse until the accelerated protons leave the plasma. The grid size is chosen as $h_x = 0.05 \lambda_L$ (direction of propagation) and $h_y = h_z = 0.4 \lambda_L$. We do, however, use a feature of VLPL that allows for the increase of cell size the further we go from the central axis in the transverse direction in order to reduce computational effort. The solver used for the simulations is the RIP solver \cite{Pukhov2020}, which requires that the time step is $\Delta t = h_x / c$.

\begin{table}
	\centering
	\caption{\label{tab:results} Results of the simulations with different ramp lengths in terms of average polarization and peak density of the proton bunch. The average polarization of the proton bunch is obtained by selecting the particles in the high-density region leaving the plasma channel. Note that for longer ramps ($75\lambda_L$, $100\lambda_L$) the shape of proton bunch is increasingly ill-defined.}
		\begin{tabular}{lccccc}
			$L_\mathrm{ramp} \; [\lambda_L]$ & 0 & 25 & 50 &75 & 100\\
			\hline
			avg. polarization $ \langle s_y \rangle$ & 0.81 & 0.83 & 0.83 & \textit{0.83} & \textit{0.63}\\
			$n_\mathrm{peak}$ [$n_\mathrm{cr}$] & 0.209 & 0.126 & 0.044 & \textit{0.027} & \textit{0.025}
			
		\end{tabular}
\end{table}

\begin{figure}
	\includegraphics[width=\textwidth]{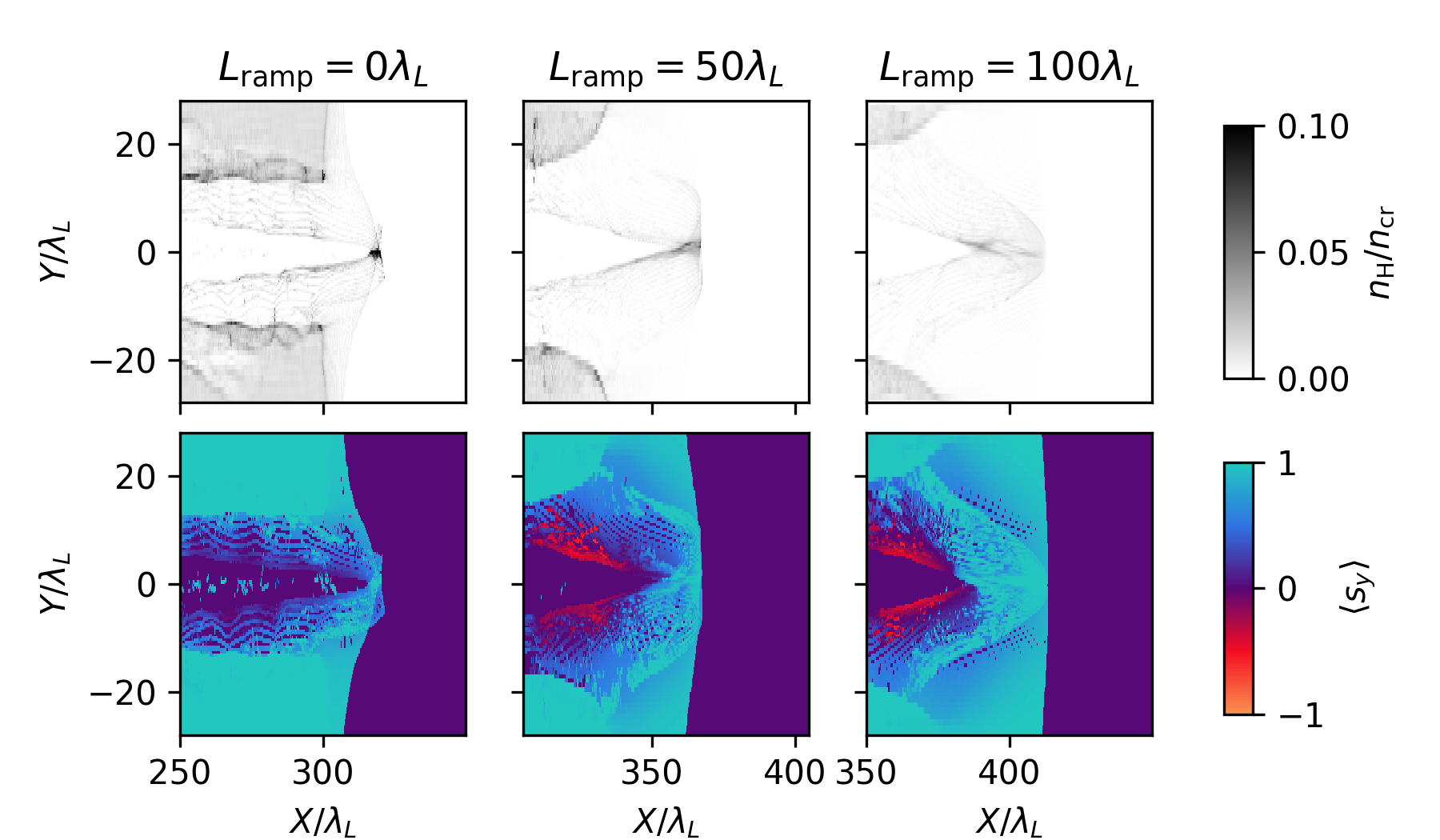}
	\caption{\label{fig:dens}Density and spin polarization for the simulations with ramp lengths $0 \lambda_L$, $50 \lambda_L$ and $100 \lambda_L$ (left to right) after the accelerated proton bunch has left the plasma (end of ramp in box middle). Note that the density plots are clipped at $0.1n_\mathrm{cr}$ for better visbility. The density plots show that increasing the ramp length is accompanied by a higher transverse spread of the resulting proton beam, which is located at $X \approx 320\lambda_L$ for the case of $L_\mathrm{ramp} = 0 \lambda_L$.}
\end{figure}

	\section{Discussion}
When the laser pulse enters the target, the electrons are driven out in the direction transverse to laser propagation, leaving behind an ionic filament that is pushed out at the end of the plasma due to the electromagnetic fields.
Since all of our simulations have the same configuration at start, the created proton bunch will be identical until the start of the down-ramp. We can see that the central filament initially keeps its polarization very well while the region around it starts to depolarize due to the electromagnetic fields (compare Fig. \ref{fig:emfields}).

As we enter the down-ramp region, we can start to see the effects of the different ramp lengths $L_\mathrm{ramp} $. For the target with a hard cut-off in density, i.e. $L_\mathrm{ramp} = 0 \lambda_L$, the usual MVA fields can be observed: the magnetic vortex starts to appear and a uniform longitudinal electric field $E_x$ that drives the protons further out of the plasma. The proton energies that can be achieved for a comparable setup are discussed in the work by Jin et al. \cite{Jin2020}, where they reached $\mathcal{E}_p \approx 53 $ MeV for a laser with $a_0 = 25$ and a HCl plasma of similar density, but with $L_\mathrm{ramp}  = 5 \lambda_L$.

\begin{figure}
	\centering
	\includegraphics[width=0.5\textwidth]{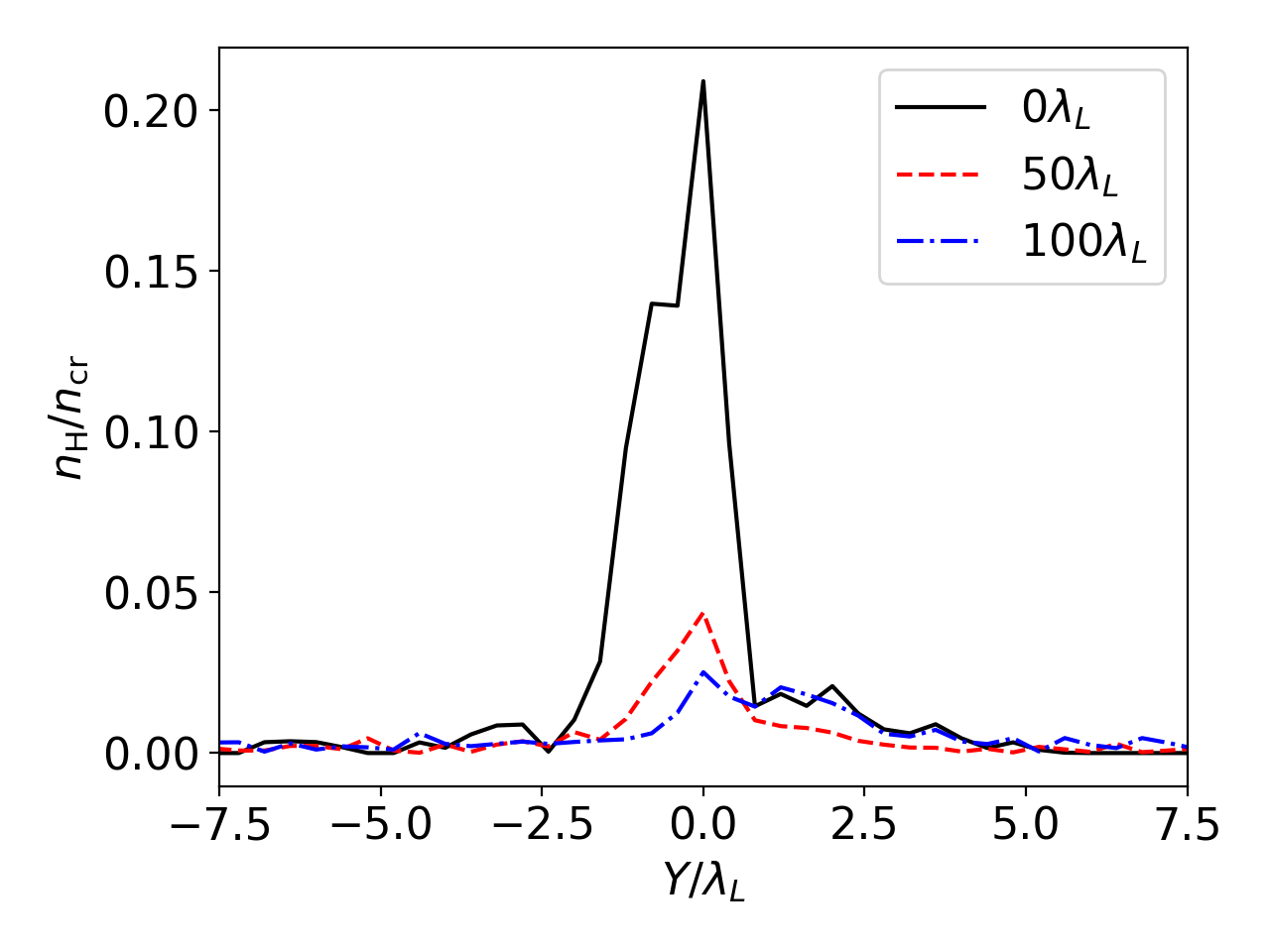}
	\caption{\label{fig:transverse}Transverse beam profile (at the plane with peak density) for a selection of different ramp lengths. Longer ramps lead to a widening of the accelerated proton beam, reducing the peak density.}
\end{figure}

Going to a longer ramp length, we can see that, due to the lower densities in those regions, the fields start to expand transversely while the proton bunch is still in the plasma (not shown here). An approximation for the strength of the magnetic field in a down-ramp is given by Nakamura et al. \cite{Nakamura2010}. This change in field configuration leads to differences clearly visible when looking the the density plots in Fig. \ref{fig:dens}: the accelerated proton bunch is modulated such that for longer ramps it further spreads in the transverse direction. Especially in the case of $L_\mathrm{ramp} = 75 \lambda_L $ and $100 \lambda_L$, the protons leaving the plasma hardly form a consistent bunch structure anymore.
Transverse density profile of the different beams as well as peak density are shown in Fig. \ref{fig:transverse} and Table \ref{tab:results}. 

\begin{figure}
	\centering
	\includegraphics[width=0.5\textwidth]{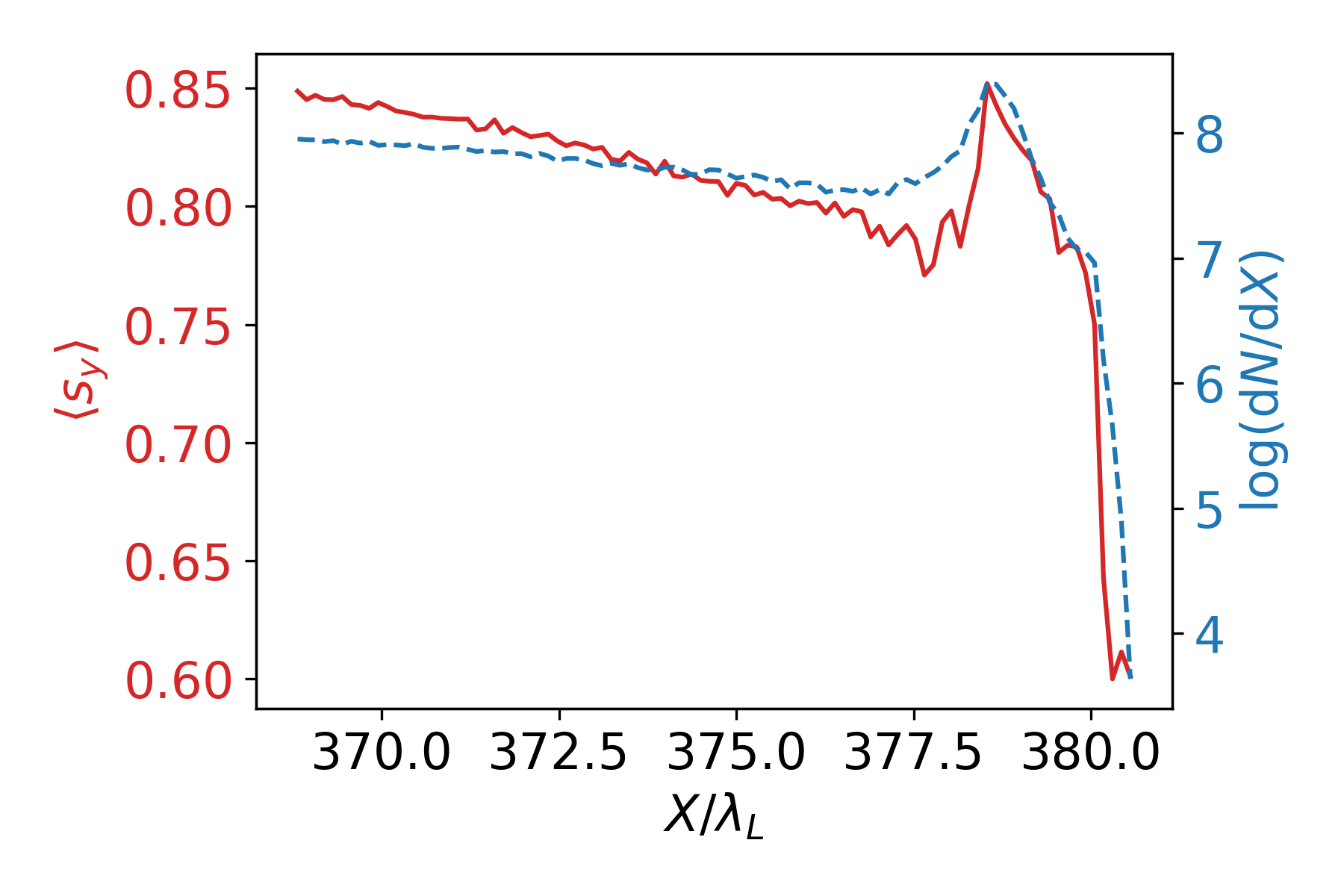}
	\caption{\label{fig:bunchpol}Exemplary polarization data for the case of $L_\mathrm{ramp} = 0 \lambda_L$ at time step $300\tau_0$. The spin for each PIC particle is assigned to a corresponding bin in the longitudinal direction for which the average spin polarization (red line) and the number of PIC particles (blue, dashed) are given.}
\end{figure}

The change in bunch structure can be attributed to two factors. Firstly, increasing the ramp lengths in a fashion as we do in our simulations, also effectively leads to a longer interaction volume, meaning that the laser is depleted of more energy. Secondly, the down-ramp allows for the transverse fields to expand, leading to a wider channel (also visible at the left boundary of the density plots in Fig. \ref{fig:dens}) and therefore the transverse growth of the proton bunch. The defocusing of the proton bunch during the passage of the down-ramp region is in agreement with the observations in \cite{Nakamura2010, Park2019}. There, the steepness of the density gradient was fixed to a value that allowed for the best collimation possible.

Besides the quality of the bunch in terms of transverse and longitudinal structure, the degree of polarization obtained at the end is of main interest.
We can directly tell by looking at the precession frequency $\Omegab$ in equation (\ref{eq:prec}) that the change in proton spin should be significantly lower than for electrons, as $|\Omegab| \propto m^{-1}$.
To measure the polarization of the bunch, we consider the particles close to the central axis. We subdivide the longitudinal direction into several bins for which we calculate the average polarization $\langle s_y \rangle$. Depending on the spatial beam structure, different degrees of polarization can be located along the volume (compare Fig. \ref{fig:bunchpol}). 
This is due to the fact that protons that end up in the beam front experience different electromagnetic fields than the ones in the beam's stern, especially when traversing through the down-ramp.

More precisely, if we subdivide our bunch into a back, middle and front part, it becomes clear that protons in the front have been focused into the bunch for the shortest amount of time. This is because here the laser pulse just has created the channel inside the plasma slab and in turn created the filament.
Therefore, the protons experience a comparatively strong field, decreasing the average polarization. In contrast, protons from the back of the bunch have been propagating for a longer period of time through the channel and consquently experiencing more depolarizing fields. This is why the spin polarization at the back of the bunch (towards $x \approx 377\lambda_L$ in Fig. \ref{fig:bunchpol}) decreases even faster than in its front (towards $x \approx 380\lambda_L$). The middle part of the bunch encounters comparatively lower field strengths and has been propagating for a moderate amount of time, yielding a higher degree of polarization than the other two parts, in accordance with the result that
\begin{equation}
	|\Omegab| \propto F := \mathrm{max}(|\Eb|, |\Bb|) \; ,
\end{equation}
which we can see from the derivation in \cite{Thomas2020}.

 As seen in Figure \ref{fig:bunchpol}, this difference in polarization between the different proton bunch ``components'' can already be seen in the absence of a down-ramp, however there it is mostly negligible as we still have a considerable average polarization.
Once we go over to longer ramps, the observations of different polarization degrees inside the bunch is strongly amplified up to the point where we see a significant reduction in average polarization. For these longer ramps, we get more lower-polarization protons since on the one hand the protons traverse through an effectively longer plasma, leading to further spin precession. On the other hand, the flatter (i.e. longer) density gradient amplifies the differences in the electromagnetic fields that the protons in the bunch's front and back experience, respectively.
This is a direct consequence of the fact that depending on the down-ramp slope, the focusing (or compression) of the bunch becomes more or less pronounced: longer ramp lengths lead to the compression of a higher amount of low polarization protons into the bunch tail (which can be seen in Fig. \ref{fig:dens}). Further, the magnetic field amplitude decreases for lower densities. Nakamura et al. \cite{Nakamura2010} have found that for a down-ramp like in our case the magnetic field decreases as

\begin{equation}
	B_2 = B_1 \frac{n_1 + n_2}{2n_1}\; ,
\end{equation}
where indices 1 and 2 denote the high- and low-density region, respectively. This gives further insight into why the front portion of the bunch has a slower decrease in polarization than the back.


In total, for most of the ramp lengths considered, a high polarization of around 80\% is maintained. Only in the simulation with $L_\mathrm{ramp} = 100 \lambda_L$ we see a significant decrease to roughly 60\%. It has, however, to be strongly emphasized that high-polarization protons are still pushed in propagation direction (see spin plot in Fig. \ref{fig:dens}), only in a non-collimated form. Instead, some protons with lower polarization (red region in the spin plots) make up a significant part of the proton bunch visible in the density plots.

These negative effects can partly be mitigated by choosing different laser-plasma parameters, although it has to be noted that for higher laser intensities the polarization degree will also decrease as it was shown in \cite{Jin2020}. This, as well as polarization decrease in case of significantly longer ramps, is explained by the scaling laws derived by Thomas et al. \cite{Thomas2020}:
A particle beam can be viewed as depolarized, once the angle between initial polarization direction and the final spin vectors is in the range of $\pi / 2$. The time after which this is to be expected is called the minimum depolarization time $T_{D,p}$ and scales as
\begin{equation}
	T_{D,p} = \frac{\pi}{6.6 a F} \; .
\end{equation}
This means that stronger electromagnetic fields induced by the laser pulse lead to a faster depolarization of the protons. Further, the longer interaction volume due to longer down-ramps also may decrease the polarization once we reach the range of the depolarization time.
It has to be noted that in the equation above, $F$ is assumed constant, i.e. this holds for constant density plasma slabs as long as the laser pulse still has most of its energy. Newly ``born'' protons, especially in down-ramp regions, can experience differing (in the ramp: lower) field strengths. 
While shorter interaction volumes are desirable for high-quality proton bunches, this may come at the cost of experimental realizability due to limitations of the nozzles and blades usable for the creation of a pre-polarized plasma target.

In a publication by A. Sharma et al. \cite{Sharma2018}, it has been shown that the ideal plasma (plateau) length for MVA scales as
\begin{equation}
	L_\mathrm{ch} = a_0 c \tau_0 \frac{n_\mathrm{cr}}{n_e} K \; ,
\end{equation}
where $K=0.074$ (in 3D) is a geometric factor. This means, that depending on the target density, we can adjust our laser parameters accordingly. Especially for lower $a_0$ we are not limited to the pulse duration we have proposed for the simulation setup. A different $\tau_0$ leads to a different time scale over which the MVA structures are built up, meaning that the collimation process of the protons into the final bunch can be aligned in such a way that we achieve both a good spatial focusing as well as the collimation of highly polarized protons.

Another option to reduce spin precession due to the prevalent electromagnetic fields would be to place a foil (e.g. made of Carbon) that is able to shield part of the fields. This setup would, however, be more in line with RPA \cite{Macchi2017}.

In the case of electrons, a mechanical setup for filtering out unwanted spin contributions has recently been proposed \cite{Wu2020}. For protons, a similar setup might be realizable.
Depolarization after the initial acceleration of the protons out of the channel gets increasingly negligible, as the prefactors of the precession frequency (\ref{eq:pref}) get smaller for higher energies $\gamma$.

Lastly, we note that to experimentally test whether longer gas-jet targets are suitable for polarized beam preparation, elements with more inert spins might be employed. It has, e.g., be shown that ${}^{129}\mathrm{Xe}$ gas can be nuclear polarized to a high degree (see \cite{Kennedy2017} and references therein). However, in this case different densities (and, consequently, laser parameters), as compared to a HCl target, have to be used.

\section{Conclusion}
We have studied the effect of down-ramp length for a near-critical HCl gas target that we use to obtain highly spin-polarized proton bunches via MVA. The interaction plasma has been pre-polarized, since polarization build-up over the course of acceleration is negligible. 
We observe that longer down-ramps modulate the spatial bunch structure, leading to ill-defined bunches. For most of the ramp lengths examined, the yielded polarization robustly stays around 80\% due to the inert proton spin.
Significantly longer ramps lead to the collimation of lower-polarization protons instead of the wanted ones. 
The difference in average bunch polarization along the propagation direction could be explained in terms of the disparate field strengths different parts of the bunch experience: the front-most part contains only recently collimated protons that therefore experience weaker fields (especially in the down-ramp). Further, they experience those fields for a shorter period of time than protons from the bunch back, which have propagated longer distances and consequently are depolarized further.
The deteriorative effects of longer down-ramps can be compensated by adjusting the parameters of the laser and plasma used to some extent. Generally, as-short-as-possible interaction volumes are preferable, since the minimum depolarization time for the bunch is inversely proportional to the field strength experienced by the protons. A next step in this subject could be an extended (semi-)analytical description of the collimation process and specifically its effect on the bunch polarization. 

\ack
	L.R. would like to thank X.F. Shen and K. Jiang for the fruitful discussions.
	This work has been supported by the DFG (project PU 213/9-1). The authors gratefully acknowledge the Gauss Centre for Supercomputing e.V. (www.gauss-centre.eu) for funding this project by providing computing time on the GCS Supercomputer JUWELS at J\"{u}lich Supercomputing Centre (JSC). The work of A.H. and M.B. has been carried out in the framework of the \textit{Ju}SPARC (J\"{u}lich Short-Pulse Particle and Radiation Center) and has been supported by the ATHENA (Accelerator Technology Helmholtz Infrastructure) consortium.

\bibliographystyle{unsrt}

\end{document}